\documentclass[sigconf]{acmart}
\AtBeginDocument{%
  }

\setcopyright{acmlicensed}
\copyrightyear{2026}
\acmYear{2026}
\setcopyright{cc}
\setcctype{by}
\acmConference[CIKM '26]{Proceedings of the 35th ACM International Conference on Information and Knowledge Management}{November 07--11, 2026}{Rome, Italy}
\acmBooktitle{Proceedings of the 35th ACM International Conference on Information and Knowledge Management (CIKM '26), November 07--11, 2026, Rome, Italy}
\acmDOI{10.1145/3799682.3840117}
\acmISBN{979-8-4007-2539-5/2026/11}

\usepackage{booktabs}
\usepackage{tabularx}
\usepackage{makecell}    
\usepackage{array}
\usepackage[ruled,vlined]{algorithm2e}
\usepackage{multirow}
\usepackage{enumitem}
\usepackage{balance}

\begin{document}


\title{LLM-based User-Product Relevance Evaluation for Dynamic Product Advertising}

\author{Wen Zhang}
\affiliation{%
  \institution{Snap Inc.}
  \city{Bellevue}
  \state{WA}
  \country{USA}
}
\email{wzhang7@snapchat.com}

\author{Xiao Bai}
\affiliation{%
  \institution{Snap Inc.}
  \city{Palo Alto}
  \state{CA}
  \country{USA}
}
\email{xbai@snapchat.com}

\author{Hsiang-wei Chao}
\affiliation{%
  \institution{Snap Inc.}
  \city{Seattle}
  \state{WA}
  \country{USA}
}
\email{hchao@snapchat.com}

\author{Mengfan Zhu}
\affiliation{%
  \institution{Snap Inc.}
  \city{Santa Monica}
  \state{CA}
  \country{USA}
}
\email{mzhu2@snapchat.com}

\author{Tong Zhao}
\affiliation{%
  \institution{Snap Inc.}
  \city{Bellevue}
  \state{WA}
  \country{USA}
}
\email{tzhao@snapchat.com}

\author{Mingxuan Ju}
\affiliation{%
  \institution{Snap Inc.}
  \city{Bellevue}
  \state{WA}
  \country{USA}
}
\email{mju@snapchat.com}

\author{Xiaodong Liu}
\affiliation{%
  \institution{Snap Inc.}
  \city{Bellevue}
  \state{WA}
  \country{USA}
}
\email{xliu9@snapchat.com}

\author{Congfei Zhang}
\affiliation{%
  \institution{Snap Inc.}
  \city{Bellevue}
  \state{WA}
  \country{USA}
}
\email{czhang3@snapchat.com}

\author{Yunzhi Zhou}
\affiliation{%
  \institution{Snap Inc.}
  \city{Palo Alto}
  \state{CA}
  \country{USA}
}
\email{yzhou10@snapchat.com}

\author{Yajun Wang}
\affiliation{%
  \institution{Snap Inc.}
  \city{Palo Alto}
  \state{CA}
  \country{USA}
}
\email{ywang30@snapchat.com}

\author{Jinchao Li}
\affiliation{%
  \institution{Snap Inc.}
  \city{Bellevue}
  \state{WA}
  \country{USA}
}
\email{jli18@snapchat.com}

\author{Yu Zhang}
\affiliation{%
  \institution{Snap Inc.}
  \city{Palo Alto}
  \state{CA}
  \country{USA}
}
\email{yzhang3@snapchat.com}

\renewcommand{\shortauthors}{Wen Zhang et al.}

\begin{abstract}
Dynamic Product Advertising (DPA) retrieval systems are a core component of modern e-commerce advertising; however, their evaluation remains challenging. Conventional offline engagement metrics are limited by sparse, biased feedback, while online evaluation is costly and time-consuming. In this work, we propose a user--product relevance metric for offline evaluation of DPA retrieval systems. The metric first compresses the engagement history into a compact set of intent representatives via Semantic-ID-grouped, event-stratified segmentation, then assesses the relevance of a recommended product to each individual intent and aggregates them as a user-level relevance score, yielding a stable and interpretable measure of retrieval quality. To compute this metric at scale, we leverage large language models (LLMs) as semantic relevance judges, capturing nuanced functional and semantic relationships that heuristic methods often miss. We propose an agentic, human-aligned prompting framework that generates multi-category relevance guidelines and distills teacher-model judgments into an efficient student model for scalable evaluation. The resulting metric strongly correlates with offline and online business metrics, enabling faster and more reliable DPA retrieval iteration before A/B testing.
\end{abstract}

\ccsdesc[500]{Information systems~Relevance assessment}


\keywords{LLM Judge, DPA, Semantic ID, Agentic, Offline Evaluation, Product Relevance, E-Commerce}


\maketitle

\vspace{-0.5em}
\section{Introduction}

Modern E-commerce advertising platforms increasingly rely on high-quality Dynamic Product Ads (DPA) to deliver personalized user experiences at scale. Within these systems, the major task of retrieval stage is efficiently filtering a massive product catalog into a small subset of candidates based on diverse user interests and engagement histories. 
DPA retrieval systems typically rely on multiple algorithms to retrieve candidate products to ensure diversity. 
Although the top-line business objective is to maximize user engagement and revenue, the industry lacks a universal and reliable metric to compare the effectiveness of different retrieval algorithms. 
Common 
offline evaluation metrics such as the recall of future engagements~\cite{zhang2025optimizing} 
are limited by selection bias and data sparsity. Unlike in the Search domain, where explicit queries provide a clear benchmark for success, the advertising domains lack a well-defined framework for user-product relevance~\cite{bauer2024exploring}. Consequently, 
machine learning practitioners often struggle to quantify whether a retrieval model is successfully surfacing relevant products or merely exploiting noisy, short-term behavioral correlations that may not lead to sustainable long-term growth.

\begin{figure*}[!tb]
  \centering
  \includegraphics[width=0.87\textwidth]{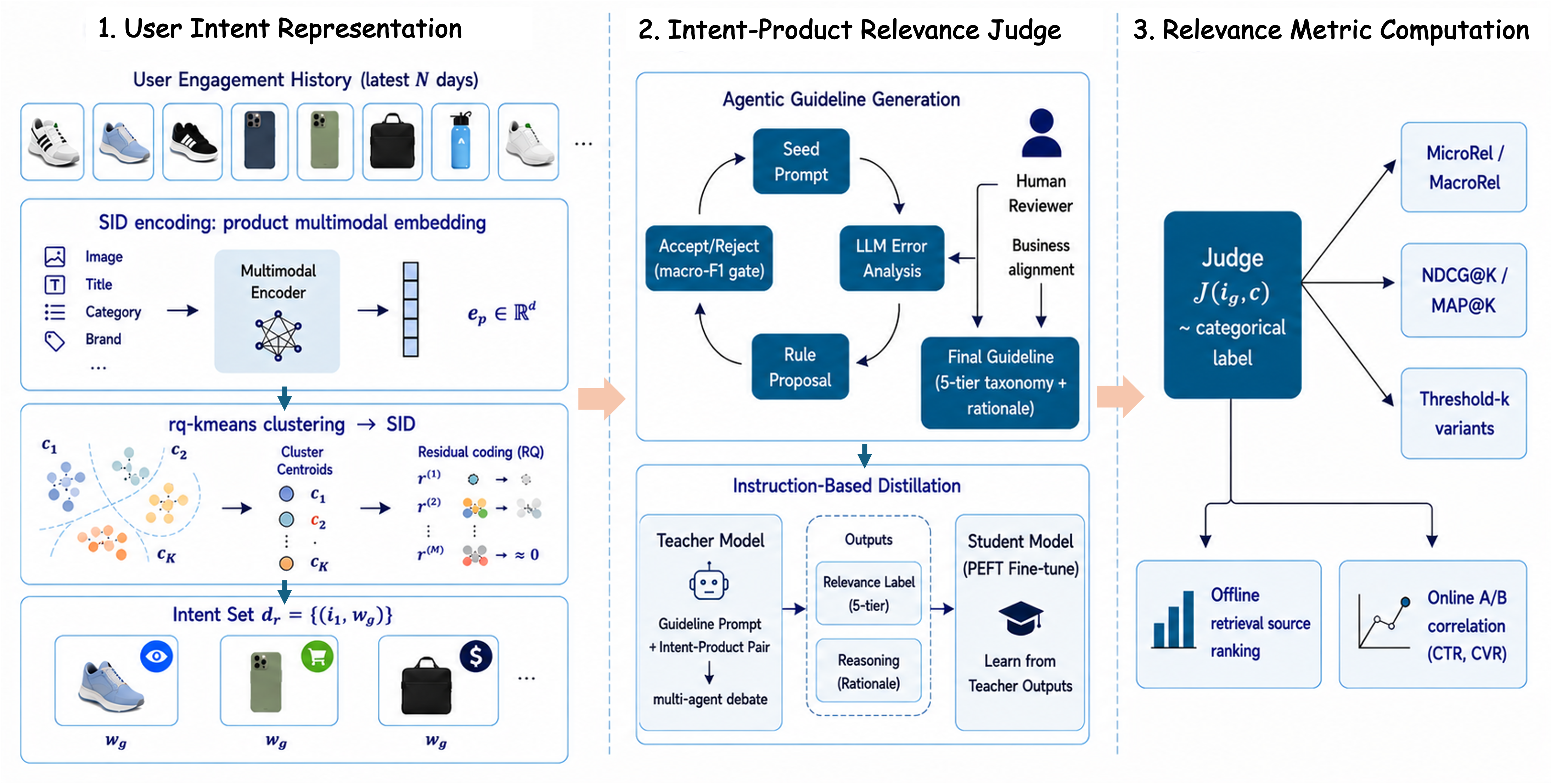}
  \vspace{-0.5em}
  \caption{LLM-based User-Product Relevance Evaluation framework. \textbf{(1) User Intent Representation}: a multimodal encoder plus residual vector quantization assigns each engaged product a hierarchical semantic ID (SID); events sharing $(\mathrm{SID}, e)$ collapse into a single weighted intent representative, yielding the compact intent set $d_r$. \textbf{(2) Intent-Product Relevance Judge}: an agentic guideline aligned with human reviewers anchors a teacher, PEFT-distilled into a student model. \textbf{(3) Metric Computation}: MicroRel/MacroRel under threshold-$k$ aggregation, validated against offline retrieval-source rankings and online CTR/CVR.}
  \Description{Diagram illustrating the stages of the LLM-based offline evaluation pipeline, including multi-agent guideline generation and student model distillation.}
  \label{fig:pipeline}
  \vspace{-0.1em}
\end{figure*} 

Defining and measuring relevance is the core obstacle. Prior work either uses explicit user studies that rate usefulness, trust, or satisfaction~\cite{pu2012evaluating, shani2015evaluating}, or implicit engagement logs of clicks, likes, shares, and reports~\cite{zhang2025optimizing}. User studies are expensive and slow~\cite{shani2015evaluating}. Engagement is abundant but noisy: clicks are biased by presentation and trust effects~\cite{joachims2017accurately}, and engagement-optimized ranking amplifies attention-grabbing content users do not actually prefer when asked~\cite{milli2025engagement}. We instead use a content-based definition of relevance: \textbf{\textit{an LLM judges each retrieved product against intents inferred from the user's engagement history, and intent-level scores are aggregated into a user-level metric.}} This reduces dependency on exposure artifacts, covers users with sparse-feedback, and makes the relevance score interpretable at the attribute level. 

Recent work uses LLMs as semantic relevance judges for search and recommendation, scoring content alignment directly rather than relying on engagement signals~\cite{sun2023chatgpt, tang2025lref, choi2025bloomintent, rahmani2025judging}. Sun et al.~\cite{sun2023chatgpt} show that prompted LLMs are effective zero-shot re-rankers and can outperform strong supervised baselines; Choi et al.~\cite{choi2025bloomintent} report substantial agreement between LLM evaluators and human raters; and large-scale benchmarks find that LLM-generated relevance labels broadly track human judgment at scale~\cite{rahmani2025judging}. These works target search-style relevance with an explicit query. In contrast, DPA retrieval system do not have search queries and typically require inferring user-level relevance signals from 
heterogeneous engagement histories.

We propose an offline evaluation framework that uses LLM judges to score user--product relevance for DPA retrieval. Two operational problems shape the design: the engagement history is heterogeneous and long, and frontier LLMs are too expensive to call on every event--candidate pair at production volume. We address both with (i) an SID (Semantic ID~\cite{singh2024better})-grouped, event-stratified compression of engagement history into a small set of weighted intent representatives, and (ii) an agentic guideline-discovery loop together with instruction-guided teacher-student finetune for production serving. Our contributions are 3-folds as follows:
\vspace{-0.5em}
\begin{itemize}[leftmargin=8pt]
    \item \textbf{Query-free relevance metric.} We define a user-level offline evaluation metric for DPA retrieval that aggregates LLM-judged intent--candidate relevance under a threshold-$k$ aggregation rule, without requiring an explicit query.
    \item \textbf{A Scalable LLM-judge Pipeline.} We propose an efficient LLM judge pipeline with 1) SID-grouped intent decomposition that reduces LLM calls by $30\times$, and 2) teacher-student distillation from an agentic, human-vetted guideline that replaces the frontier teacher with an efficient student at production cost.
    \item \textbf{Online validation.} The metric ranks six retrieval sources consistently with conversion precision/recall - how well the retrieved candidates aligns with user's subsequence purchases - and is directionally aligned with online CTR/CVR deltas across four A/B tests; It thus provides a reliable view of retrieval system health and a low-cost signal for guiding offline model iteration.
\end{itemize}

\vspace{-0.5em}
\section{LLM-Based User-Product Relevance Metric}
The proposed relevance evaluation framework has three blocks (Figure~\ref{fig:pipeline}): a compact \emph{user intent representation} that segments engagement history into SID-grouped representatives (§\ref{sec:intent}); an \emph{intent-product relevance judge} built from an agentically-discovered guideline (§\ref{sec:guideline}) and a teacher-student distillation (§\ref{sec:student}); and a \emph{metric-computation layer} (§\ref{sec:metric}) that aggregates pairwise judgments into request- and user-level scores.

\subsection{User Intent Representation via SID-based Segmentation} \label{sec:intent}
A user is represented by her engagement history $H_u = \{(p_t, e_t)\}_{t=1}^{T}$, where $p_t$ is an engaged product and $e_t$ is the event type (view, add-to-cart, purchase, etc.). Treating every event as a distinct intent is the most expressive representation but costly in production: a typical $30$-day window holds hundreds of events, most of which are semantically redundant, inflating LLM cost and biasing the metric toward retrieval algorithms that re-surface near-duplicates of frequently-viewed items.

\paragraph{SID grouping.}
To select a compact, signal-rich representative set, we run a lightweight grouping step before the judge. Each product in the catalog is mapped to a hierarchical \emph{Semantic ID} (SID) via k-means residual vector quantization (RQ-kmeans) over an in-house content-based multimodal product embedding \cite{deng2025onerec}, using $L$ stages with $512$-entry codebooks to provide sufficient granularity. Concretely, with $r_0\!=\!\mathrm{embed}(p)$ and $C_l\!\in\!\mathbb{R}^{512\times d}$ the level-$l$ codebook,
\vspace{-0.3em}
\begin{equation}
\small
\mathrm{SID}(p) = (c_1,\ldots,c_L), \;\; c_l = \arg\min_k \|r_{l-1} - C_l^{(k)}\|_2, \;\; r_l = r_{l-1} - C_l^{(c_l)},
\label{eq:sid}
\end{equation}
so each stage encodes the residual left by the previous one and longer prefixes refine coarser ones. The length of the leading prefix is used as the grouping key to control granularity: a length-$1$ prefix groups products at roughly category-prototype resolution (e.g., all running shoes), while the full length-$L$ prefix distinguishes near product-type-plus-attribute granularity (e.g., a specific brand-model-color combination). Two items sharing the chosen-length prefix are treated as content-equivalent; The SID lookup is precomputed and cached, so its runtime cost is negligible during LLM judge inference. With $L\!=\!3$ in production -- a balance between cost (fewer LLM calls at shallow $L$) and granularity (longer prefixes preserve product-level distinctions) -- we partition $H_u$ by $(\mathrm{SID}(p_t), e_t)$ and elect one representative $i_g$ per group -- the most recent product, with ties broken by event-type strength (e.g., \textit{purchase} $>$ \textit{add-to-cart} $>$ \textit{others}). The resulting weighted intent set is
\vspace{-0.3em}
\begin{equation}
\small
d_u = \{(i_g, w_g) : g \in \mathrm{Groups}(H_u)\}, \;\; w_g = \log(1+|g|)\cdot\beta(e_g),
\label{eq:intent_set}
\end{equation}
with $\beta(\text{purchase})\!=\!3$, $\beta(\text{add-to-cart})\!=\!2$, $\beta(\text{others})\!=\!1$. The $\log(1+|g|)$ factor lets a heavily-engaged content-equivalence group dominate over a one-off view of an unrelated product without scaling linearly with view count; event-type stratification keeps a \textit{low-intent} signal from overruling a same-SID \textit{high-intent} signal.

\vspace{-0.5em}
\subsection{Intent-Product Relevance Judge} \label{sec:judge}
Given the intent set $d_r$ from §\ref{sec:intent}, we need a judge $\mathcal{J}(i, c)$ that assigns a relevance label to each (intent representative $i$, candidate product $c$). Two issues shape the design. The taxonomy must capture DPA-specific product semantics without becoming so detailed that a frontier LLM applies it inconsistently; and running a frontier LLM as $\mathcal{J}$ at production volume is too expensive. We address the first with an agentic guideline-discovery loop followed by a human-alignment pass (§\ref{sec:guideline}), and the second with knowledge distillation into a small student (§\ref{sec:student}).

\vspace{-0.1em}
\subsubsection{Agentic Guideline Generation} \label{sec:guideline}
\label{llm_guideline}
We build the evaluation guideline in two phases. Phase A is an \emph{agentic loop} where an LLM judge iteratively refines its own decision rules from labeled examples; Phase B is a \emph{human alignment} pass where editors steer the resulting rule set toward DPA business priorities.

\paragraph{Phase A: agentic rule discovery.}
We start from a minimal \emph{criteria-only} seed prompt -- a single instruction, ``\textit{Label the candidate as Exact, Similar, Substitute, Complementary, or Irrelevant}'' -- with no per-class definitions, examples, or attribute hints. A frontier LLM (e.g., Gemini-2.5-Pro) labels a stratified calibration split using this seed prompt, scored by per-class F1 against the human labels. The refinement loop iterates (i) clustering mis-classified pairs by the (predicted, label) confusion and the judge's free-form rationale; (ii) prompting a separate reasoner to propose up to three new sub-rules targeting the true-label class for the largest cluster; and (iii) provisionally appending each sub-rule, re-labeling the calibration set, and accepting iff the paired-bootstrap 95\% CI on $\Delta$accuracy excludes zero with no per-class F1 regression beyond $0.5$ percentage points. Rejected sub-rules are logged as ``previously tried''. A round ends on the first accepted sub-rule or after all proposals are rejected; the loop stops after $N$ rounds or two consecutive empty rounds. We run multiple trials and retain only sub-rules that recur across trials, stabilizing the discovered set against proposer stochasticity. The candidate-rule pool surfaces seven of the twelve editorial-guideline reasons (attribute-level: type, brand, model/style, size, color, material, component-vs-whole); the five not surfaced -- demographic gender/age, broad-Complementary/Irrelevant, asymmetric Complementary -- are deferred to Phase B.

\paragraph{Phase B: human business-alignment.}
The agentic phase optimizes calibration-set agreement but is silent on downstream business goals. Human reviewers audit the Phase-A output against three criteria: coverage of business-critical mismatches (demographic gender/age, regulated categories, advertiser exclusions), editorial-standard compatibility, and reproducibility on a $200$-pair business-stratified sample. Vetoed rules return to Phase A with an explanatory note. The most consequential intervention in our deployment promoted demographic mismatch from a \textit{Substitute}-leaning signal to a hard \textit{Irrelevant} override -- a rule the loop repeatedly deprioritized because demographic-mismatch pairs are statistically rare in the calibration set. Phase B is load-bearing: demographic-override rules emerge from editorial review rather than from the agentic loop, because demographic-mismatch pairs are too rare in the calibration set to drive the loop's statistical gate.

\paragraph{Final taxonomy and rationale format.}
The resulting guideline -- the seed taxonomy refined by Phase-A sub-rules under the Phase-B business overrides -- is what the rest of the paper refers to as ``the guideline''. It instructs the LLM judge to assign each (intent, candidate) pair to one of five tiers and to emit a structured rationale by selecting from a controlled attribute vocabulary (\textit{Product Type}, \textit{Brand}, \textit{Demographics}, \textit{Component-vs-Whole}, etc.). The final taxonomy summarization:

\begin{itemize}[leftmargin=8pt,itemsep=0pt,topsep=2pt]
    \item \textbf{Exact:} matches all specific attributes of the intent (type, brand, model, size, color).
    \item \textbf{Similar:} same product category, differs in brand, model, or color.
    \item \textbf{Substitute:} different type but functionally interchangeable.
    \item \textbf{Complementary:} different type, typically used together with the intent~\cite{jeon2025leveraging}.
    \item \textbf{Irrelevant:} no functional or categorical link, or a critical demographic mismatch.
\end{itemize}


\vspace{-0.5em}
\subsubsection{Teacher-Student Knowledge Distillation} \label{sec:student}
\label{student_llm}
The finalized guideline from §\ref{sec:guideline} could in principle be deployed as an instruction prompt to a frontier LLM at serving time, but at production volume this is prohibitively expensive (Table~\ref{tab:efficiency}). We instead use the guideline-prompted frontier LLM as a \emph{teacher} that emits high-quality labels and rationales on a curated training corpus, then fine-tune a much smaller \emph{student} LLM on those labels. The student is what we deploy as $\mathcal{J}$.

\paragraph{Distillation dataset construction.}
We construct a large distillation corpus that reflects real-world retrieval scenarios via a four-stage pipeline. \textbf{Stratified seed sampling:} intent representatives $i_g$ are sampled stratified across user engagements and demographics (age, gender, engagement type, engagement frequency), ensuring broader and more balanced pattern coverage than a uniform random distribution. \textbf{Multi-source candidate generation:} for each seed, candidates are aggregated from three complementary streams -- production retrieval logs, historical conversion-driven engagement signals, and curated high-value advertiser groups. \textbf{Hard-case mining:} we downsample to $N\!=\!10$ candidates per seed using a multimodal embedding filter to intentionally surface semantically-similar-but-irrelevant negatives. \textbf{Labeling:} pairs are labeled by the teacher protocol above; rationale traces are retained as auxiliary supervision for the student.

\paragraph{Teacher labeling.}
The finalized guideline is loaded as the instruction prompt to a frontier LLM (Gemini-2.5-Pro), which annotates each seed--candidate pair in the distillation datasets with a relation label and an attribute-level rationale. GPT-5 runs in parallel under the same guideline as a verification agent; pairs on which the two agents agree at high self-reported confidence are accepted directly, and the disagreed cases are sent to human auditors for relabeling.

\paragraph{Student fine-tuning.}
We test both Gemini-2.5-Flash and Gemini-2.5-Flash-Lite as student backbones and apply parameter-efficient instruction fine-tuning (LoRA with adapter rank $16$, learning-rate multiplier $2$, $5$ epochs). The student is trained for multi-label classification over the five-tier taxonomy and jointly emits an attribute-level rationale conditioned on the label, matching the teacher's output format. At serving time, the student replaces both teacher LLMs in the inference path; the teacher is invoked only on guideline-update or distillation-refresh cycles.

\vspace{-0.5em}
\subsection{Metric Computation} \label{sec:metric}
Given the intent set $d_u$ from §\ref{sec:intent} and the pairwise judge $\mathcal{J}$ from §\ref{sec:judge}, we compute a per-user relevance score $s(u)$ and then aggregate across users into a single retrieval-quality metric. The aggregation layer is decoupled from the upstream blocks so the same judgments can be re-aggregated under different rules without re-invoking the LLM. Algorithm~\ref{alg:offline_metric} summarizes the procedure.

\begin{algorithm}[t]
\caption{Offline User--Product Relevance Metric}
\label{alg:offline_metric}
\small
\KwIn{A group of users $G$ (e.g., users routed to a specific retrieval source); per-user engagement history $H_u{=}\{(p_t, e_t)\}$ and candidate set $C_u$ for each $u\in G$; SID truncation $L$; event-type weights $\beta(\cdot)$; threshold $k$; judge $\mathcal{J}(i, c)$.}
\KwOut{Per-user scores $\{s(u)\}_{u \in G}$; group-level summaries $\text{MicroRel}(G)$, $\text{MacroRel}(G)$.}

\tcp{\footnotesize Step 1: per-user intent set + relevance score}
\ForEach{$u \in G$}{
    $\mathcal{P}_u \gets \texttt{Partition}\big(H_u,\ \text{key}{=}(\mathrm{SID}_{L}(p_t),\, e_t)\big)$ \;
    $d_u \gets \big\{\, (i_g,\, w_g) \,:\, g \in \mathcal{P}_u \,\big\}$, where $i_g\!=\!\arg\max_{(p,e)\in g} \mathrm{recency}(p)$,\, $w_g\!=\!\log(1{+}|g|)\!\cdot\!\beta(e_g)$ \;
    \ForEach{$c \in C_u$}{
        $m_u(c) \gets \sum_{(i_g, w_g) \in d_u} w_g \cdot \mathcal{J}(i_g, c)$ \tcp*[r]{weighted intent-coverage}
        $v_u(c) \gets \mathbb{I}[\,m_u(c) \ge k\,]$ \tcp*[r]{candidate-level binary relevance}
    }
    $s(u) \gets \tfrac{1}{|C_u|} \sum_{c \in C_u} v_u(c)$ \tcp*[r]{per-user score}
}
\tcp{\footnotesize Step 2: group-level aggregation}
$\text{MicroRel}(G) \gets \dfrac{\sum_{u \in G} \sum_{c \in C_u} v_u(c)}{\sum_{u \in G} |C_u|}$ \tcp*[r]{$(u,c)$-equal weighting}
$\text{MacroRel}(G) \gets \dfrac{1}{|G|} \sum_{u \in G} s(u)$ \tcp*[r]{user-equal weighting}
\Return{$\{s(u)\}_{u \in G},\, \text{MicroRel}(G), \text{MacroRel}(G)$}\;
\end{algorithm}

\paragraph{Per-user score.}
For each user $u$ and candidate $c \in C_u$, the judge $\mathcal{J}$ emits a categorical label that we map to a per-pair score (binary $\{0,1\}$ by default; alternatively a 5-grade gain $g \in \{0,1,2,3,4\}$ following~\cite{reddy2022shopping}). The weighted intent-coverage $m_u(c) = \sum_{(i_g, w_g) \in d_u} w_g \cdot \mathcal{J}(i_g, c)$ aggregates the per-pair signal across the intent representatives, and the threshold $k$ converts it to a binary candidate-level relevance $v_u(c) = \mathbb{I}[m_u(c) \ge k]$. The per-user score $s(u) = \frac{1}{|C_u|}\sum_{c \in C_u} v_u(c)$ is the primary output of the metric layer.

\paragraph{User vs.\ group aggregation.}
The score $s(u)$ can be reported per user, or aggregated to any group $G$ of users (retrieval source, demographic bucket, etc.). We use two complementary summaries: \textbf{MicroRel}$(G) = \sum_{u \in G}\sum_c v_u(c) \,/\, \sum_{u \in G} |C_u|$ weights each $(u, c)$ pair equally; \textbf{MacroRel}$(G) = \frac{1}{|G|}\sum_{u \in G} s(u)$ weights each user equally and is therefore less sensitive to skewed traffic. In the rest of the paper we group by retrieval source; the same machinery applies to any other user grouping.

\paragraph{Threshold $k$ as a strictness knob.}
We operate at $k\!=\!7$ in production (justified in §\ref{sec:abl-threshold-k}); the simpler $k\!=\!1$ at-least-one rule is included as a permissive baseline since it equates a single weak intent match with broad intent coverage.

\vspace{-0.5em}
\section{Online-Offline Retrieval Metrics Alignment}
In this section, we benchmark the proposed LLM-based user-product relevance metric against conventional DPA retrieval offline metrics (e.g., \textbf{\textit{Conversion Precision/Recall}}~\cite{zhang2025optimizing}) and evaluate their consistency across various retrieval sources. A key advantage of the LLM-based approach is its independence from online traffic data. While conventional metrics often rely on future engagement signals to accurately measure precision/recall, our metric operates strictly offline, providing a high-fidelity relevance signal without the need for real-time interaction logs.

\vspace{-0.5em}
\subsection{Online Conversion Precision/Recall}

Utilizing online traffic, we employ a look-back evaluation strategy that links observed conversions to their contributing ad requests ~\cite{zhang2025optimizing}. We define relevance using a strict exact-match criterion, where a retrieved product must be identical to the converted item within a one-day post-impression attribution window. This focus assesses the system's ability to surface the precise product driving a transaction. Accordingly, Conversion Precision (\textbf{CP}) is the proportion of retrieved products that match the conversion, while Conversion Recall (\textbf{CR}) measures the fraction of ultimately converted products successfully surfaced by the system. Metrics are limited to users with both an eligible ad request and a subsequent confirmed conversion.




\subsection{Retrieval Source Baselines} \label{sec:baselines}
A production DPA retrieval system blends several heterogeneous sources, each with distinct algorithmic and signal-coverage profiles, so a single offline metric that ranks them consistently must generalize across signal types. To make our evaluation reproducible and to expose where each source is expected to win or lose, we describe the six representative retrieval sources used throughout this study.
\textbf{Matched Features (MF)} is a deterministic retargeting source: candidates are products whose canonical IDs appear in the user's recent engagement log, optionally expanded by the seller's product-graph siblings. Coverage is narrow but precision is high by construction.
\textbf{Similar User (SU)} is a two-tower collaborative model trained on $(u, p)$ engagement pairs with sampled softmax; at serving time the user tower's vector is used to retrieve the top items from an ANN index over the global active catalog.
\textbf{Content-Based Retrieval (CBR)} encodes each catalog item with an in-house multimodal encoder (image + title + structured attributes) and serves products that are nearest neighbors of the user's recently engaged items in this content embedding space.
\textbf{Engagement-Based Retrieval (EBR)} is a sequence-aware retrieval tower fine-tuned on impression-attributed engagement; it shares the same ANN backend as CBR but its encoder is trained on behavioral co-occurrence rather than content similarity, so it captures complementary patterns such as cross-category purchase pairs.
\textbf{Popularity-Based Retrieval (PBR)} ranks the global catalog by a smoothed conversion rate over the trailing $N$d, segmented by coarse demographic buckets; it has no per-user signal beyond the segment.
The \textbf{Random Baseline (Rand)} samples products uniformly from a coarse category bucket aligned with the user's last engaged category. All sources retrieve a top-$K{=}15$ candidate set per request; the same downstream ranker and budget policy is applied so observed differences are attributable to the retrieval stage.

\begin{table}[t]
\centering
\footnotesize
\setlength{\tabcolsep}{2.5pt}
\caption{Retrieval source performance, reported as \textit{log-rel} against the Random Baseline, i.e., $\log_{10}(\text{method}/\text{Rand})$ measures relative improvement (higher is better). And () indicates the ranking across all retrieval sources. }
\label{tab:retrieval_sources_log_relative}
\begin{tabular}{lccccccc}
\toprule
\textbf{Src} & \textbf{CP} & \textbf{CR} & \textbf{NDCG@15} & \textbf{MAP@15} & \textbf{MicroRel} & \textbf{MacroRel} & \textbf{ImpRel} \\
\midrule
MF       & +3.13 (1) & +0.91 (1) & +0.50 (1) & +0.29 (1) & +1.21 (1) & +1.02 (1) & +0.60 (1) \\
SU       & +2.04 (2) & +0.37 (2) & +0.16 (4) & +0.19 (2) & +0.75 (2) & +0.30 (2) & +0.39 (2) \\
PBR      & +0.45 (3) & +0.59 (3) & -0.03 (6) & +0.03 (4) & +0.12 (5) & -0.08 (6) & +0.18 (4) \\
CBR      & +0.15 (4) & +0.30 (4) & +0.35 (2) & +0.11 (3) & +0.32 (3) & +0.07 (3) & +0.23 (3) \\
EBR      & +0.15 (5) & +0.29 (5) & +0.28 (3) & +0.01 (5) & +0.24 (4) & +0.05 (4) & +0.12 (5) \\
Rand     & 0.00 (6)  & 0.00 (6)  & 0.00 (5)  & 0.00 (6)  & 0.00 (6)  & 0.00 (5)  & 0.00 (6) \\
\bottomrule
\end{tabular}
\vspace{-0.5em}
\end{table}

\subsection{Alignment Between User-Product Relevance and Conversion-Based Metrics}
Table~\ref{tab:retrieval_sources_log_relative} reports the LLM-based offline metrics alongside conversion precision/recall (CP, CR) over a $30$-day production window. NDCG@15 and MAP@15 are rank-aware retrieval metrics computed on the same LLM-judge labels, while the remaining columns follow the convention in §\ref{sec:metric}. The two sets of rankings agree closely, indicating that the LLM judge tracks conversion-based outcomes.
We also report $ImpRel$, a metric derived from impressions selected by the downstream click prediction model from the initial retrieval pool, which captures end-to-end performance. The rankings mostly agree: CP and MicroRel produce the same ordering except for Popularity-Based Retrieval, which carries global engagement trends rather than semantic content correlation. Pearson correlations between MicroRel and CP/CR/CVR under the $k\!=\!7$ aggregation default are reported in Figure~\ref{fig:threshold_k}. Across the six retrieval sources, MicroRel achieves the highest cross-source Pearson with CP ($r\!=\!0.97$), with ImpRel ($r\!=\!0.95$) and MacroRel ($r\!=\!0.93$) close behind. Standard rank-aware retrieval metrics computed on the same LLM-judge labels are weaker proxies for the production signal: MAP@15 reaches $r\!=\!0.94$ and NDCG@15 only $r\!=\!0.55$. Intent-aware aggregation therefore tracks online conversion better than off-the-shelf rank-aware metrics. 

\begin{table}[t]
\centering
\footnotesize
\setlength{\tabcolsep}{3pt}
\caption{Offline vs.\ online metrics for two retrieval families (content-based, engagement-based) across four online experiments (two for each family). log-rel reports each version against its controls, so positive values across both rows reflect successive improvements. All four online CTR/CVR deltas are statistically significant under our internal A/B framework within 14 days of windows with 10\% of DPA traffics.}
\label{tab:online_retrieval_metrics}
\begin{tabularx}{\columnwidth}{X c c c c}
\toprule
Retrieval Source 
& \multicolumn{2}{c}{Offline LLM Judge} 
& \multicolumn{2}{c}{Online A/B Metrics} \\
\cmidrule(lr){2-3} \cmidrule(lr){4-5}
& MicroRel & MacroRel & CTR & CVR \\
\midrule
Content-based Retrieval Model 1     & +0.138 & +0.122 & +0.149\% & +1.472\% \\
Content-based Retrieval Model 2     & +0.176 & +0.169 & +0.211\% & +4.98\%  \\
\midrule
Engagement-based Retrieval Model 1  & +0.068 & +0.019 & +0.350\% & +2.604\% \\
Engagement-based Retrieval Model 2  & +0.062 & +0.095 & +0.210\% & +2.194\% \\
\bottomrule
\end{tabularx}
\end{table}

\subsection{Alignment Between User-Product Relevance and Online Metrics}
A natural follow-up: \textbf{\textit{do offline gains in the relevance metric translate into online wins?}} Table~\ref{tab:online_retrieval_metrics} shows directional agreement between offline relevance improvements and online CTR/CVR gains, so the offline metric is usable for guiding DPA retrieval model development before paying for an A/B test. In this experiment, we analyze two primary retrieval categories within our DPA system: content-based and engagement-based retrieval. We evaluate the LLM-based relevance scores across several iterations of these models to assess performance improvements. 
We then further validate them through online A/B experiments in the DPA online system. In Table~\ref{tab:online_retrieval_metrics}, Model~1 and Model~2 are two successive iterations of each retrieval family; each row reports its version against its prior baseline (Model~1 vs.\ the preceding production model; Model~2 vs.\ Model~1) using $\text{log-rel}=\log_{10}(\text{NewVersionRelevance}/\text{BaseVersionRelevance})$, so positive values consistently indicate improvement.
Interestingly, engagement-based retrieval exhibits relatively weak correlations, likely because the enhanced capability reflects a blend of engagement signals and content features.

\vspace{-0.5em}
\paragraph{What changed across iterations.}
The two retrieval families were iterated along orthogonal axes; within each family, the same axes apply to both Model~1 (over the preceding production model) and Model~2 (over Model~1). Each content-based iteration enlarges the training corpus to broaden catalog coverage and refines the seed-product selection strategy used to construct (seed, candidate) training pairs, together improving content alignment for under-represented product types. Each engagement-based iteration strengthens alignment with the downstream product ranker by jointly tuning their objectives and incorporates more engagement signals directly correlated with CVR (e.g., upweighting conversion-attributed events). Because the engagement-based improvement mixes ranker-alignment with the underlying engagement signal, its offline-online delta is smaller and noisier, which explains the weaker correlation in Table~\ref{tab:online_retrieval_metrics}.



\begin{table}[t]
\centering
\footnotesize
\setlength{\tabcolsep}{3pt}
\caption{Intent-set construction ablation at $k\!=\!7$. Column definitions: see §\ref{sec:abl-intent}.}
\vspace{-0.5em}
\label{tab:sid_ablation}
\begin{tabular}{lccccc}
\toprule
\textbf{Intent-set Variant} & \textbf{Cost} & \textbf{Spear.\,vs.\,Full} & \textbf{P(CVR)} & \textbf{P(CR)} & \textbf{P(CP)} \\
\midrule
Full sequence (baseline)                       & $100\%$  & $1.000$          & $0.24$ & $0.19$ & $0.32$ \\
Random subsample ($25\%$ L\_seq)           & $25\%$   & $0.905$          & $0.26$ & $0.23$ & $0.19$ \\
Last 1-day window                              & $11\%$   & $0.528$          & $0.21$ & $0.25$ & $0.31$ \\
SID only, no event split                       & $3.4\%$  & $0.990$          & $0.30$ & $0.17$ & $0.24$ \\
SID + event split, uniform                     & $3.4\%$  & $\mathbf{0.991}$ & $0.29$ & $0.18$ & $0.26$ \\
\textbf{SID + event split, weighted}    & $\mathbf{3.4\%}$ & $0.989$  & $\mathbf{0.50}$ & $\mathbf{0.29}$ & $\mathbf{0.35}$ \\
\bottomrule
\end{tabular}
\vspace{-0.5em}
\end{table}

\section{Ablation Study}
We construct a 10K dataset from retrieval logs, audited through multiple rounds of human review to reach consensus. To ensure reliable labels, we adopt a multi-agent framework in which GPT-5 and Gemini-2.5-Pro independently generate judgments that are aggregated via confidence scores; human annotators resolve disagreements until convergence, and the final judgments are treated as ground truth. Results of the main ablation are shown in Table~\ref{tab:model_vs_human_metrics}. In addition, we ablate two design choices that are central to the framework but whose empirical justification would have interrupted the methodology in §\ref{sec:intent} and §\ref{sec:guideline}: the SID-based intent-set construction (§\ref{sec:abl-intent}), and the two-phase agentic-then-human guideline-generation procedure (§\ref{sec:abl-guidelines}).

\vspace{-0.5em}
\subsection{Intent-set construction} \label{sec:abl-intent}
We ablate the intent-set decomposition along three axes -- (i) random vs.\ recency-based vs.\ SID-based selection, (ii) presence or absence of event-type stratification, and (iii) uniform vs.\ size-and-event-weighted aggregation -- to isolate the contribution of each design choice rather than of subsampling alone. To put all variants on the same label matrix, we exhaustively judge every event--candidate pair in the $30$-day window with the teacher LLM, then for each variant compute its offline MicroRel and report the Pearson correlation against per-(date, source) live CVR, CR, and CP. In Table~\ref{tab:sid_ablation}, \emph{Cost} is pairwise calls as a percentage of the raw $30$-day history (several hundred events per user after product-id deduplication), placing all variants on a common denominator; \emph{Spear.\,vs.\,Full} is the per-user MicroRel rank correlation against the full sequence; and $P(\cdot)$ is the Pearson between offline MicroRel and the named online metric. Random baselines average over multiple seeds.

\begin{table}[t]
\centering
\small
\setlength{\tabcolsep}{3pt}
\caption{Impact of Guidelines and Student Models}
\vspace{-0.5em}
\label{tab:model_vs_human_metrics}
\begin{tabular}{llccc}
\toprule
\textbf{Role} & \textbf{Model Config} & \textbf{Precision} & \textbf{Recall} & \textbf{F1} \\
\midrule
\multirow{2}{*}{Teacher}
 & Gemini-2.5-Pro (no guidel.) & 0.824 & 0.825 & 0.816 \\
 & Gemini-2.5-Pro (with guidel.) & 0.861 & 0.866 & 0.860 \\
\midrule
\multirow{4}{*}{Student}
 & SigLIP-2-mm~\cite{tschannen2025siglip}            & 0.780 & 0.751 & 0.746 \\
 & Gemini-2.5-Flash (with guidel.) & 0.840 & 0.843 & 0.832 \\
 & Gemini-2.5-Flash-ft         & \textbf{0.869} & \textbf{0.873} & \textbf{0.869} \\
 & Gemini-2.5-Flash-lite-ft    & 0.845 & 0.849 & 0.846 \\
\bottomrule
\end{tabular}
\vspace{-0.5em}
\end{table}

Two findings stand out. First, recency truncation halves intent-set fidelity (Last 1-day Spearman $0.53$ vs.\ Full's $1.00$) and yet barely moves online Pearson -- a Last 1-day intent set tracks CP/CR/CVR within $\pm 0.03$--$0.06$ of the Full sequence. The signal that drives online business metrics is concentrated in a small, recent slice of engagement, even though the intent-set structure looks very different. This is what makes aggressive compression viable in the first place. Second, SID grouping with event-type stratification keeps that compression while also restoring the intent-set fidelity: it matches the full sequence at $0.99$ Spearman at only $3.4\%$ of its pairwise cost -- a $\sim$$30\times$ reduction in LLM calls -- and the explicit $w_g{=}\log(1{+}|g|)\cdot\beta(e_g)$ weighting lifts the variant to P(CVR)$\,=\,0.50$, P(CR)$\,=\,0.29$, P(CP)$\,=\,0.35$, beating Full and every other variant on all three online metrics. The margin over the Random subsample baseline (which itself costs $7\times$ more) is $+0.24$ on CVR and $+0.16$ on CP. We attribute this to a regularization effect: by deduplicating near-identical events and explicitly downweighting low-intent views, the metric stops rewarding retrieval algorithms that re-surface repeated views of the same item -- a failure mode the full-sequence MicroRel cannot distinguish from genuine relevance.

\vspace{-0.5em}
\subsection{Impact of Guidelines} \label{sec:abl-guidelines}
Using Gemini-2.5-Pro as the backbone, a judge given only brief relevance definitions (without the guideline) is 4\% worse on both precision and recall than the guideline-prompted version (Table~\ref{tab:model_vs_human_metrics}) and even underperforms the smaller student model with guideline, showing that the guideline carries most of the lift. The gain concentrates on rule classes that the loose prompt systematically misjudges, e.g., demographic mismatches and brand-vs.-category confusions, which the agentic discovery loop in §\ref{sec:guideline} surfaces. 

\vspace{-0.5em}
\subsection{Choice of Student Judge} \label{sec:abl-student}
We adopt Gemini-2.5-Flash as the student backbone. Instruction fine-tuning improves precision by 2.9\% and recall by 3.0\%, outperforming detailed prompting alone and fine-tuned Gemini-2.5-Flash-Lite. The fine-tuned Flash student in fact matches the Pro teacher on $F_1$ ($0.869$ vs.\ $0.860$) while running at a fraction of the cost (Table~\ref{tab:efficiency}), so we deploy it as the production judge $\mathcal{J}$. It surpasses SigLIP2-MM---which uses 10$\times$ more training data and a two-layer classifier over fused intent--product embeddings---by 9\% in precision and 12\% in recall. The lightweight classifier struggles with complex relevance patterns learning from limited training data, highlighting the advantage of LLM-based judges with few-shot learning.

\begin{figure}[t]
\centering
\includegraphics[width=0.9\columnwidth]{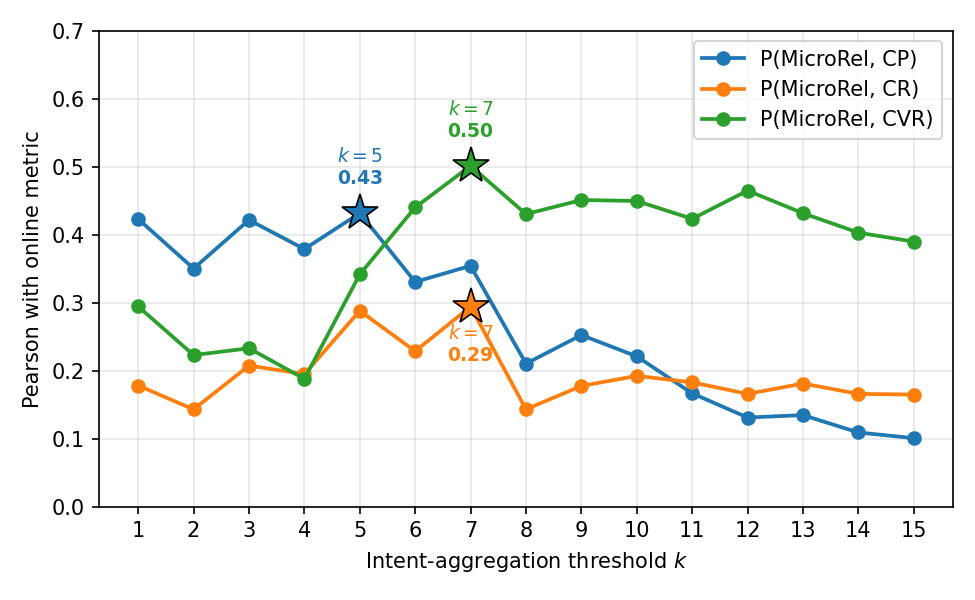}
\vspace{-0.5em}
\caption{MicroRel's Pearson correlation with live CP, CR, and CVR as a function of the intent-aggregation threshold $k$. Stars mark per-metric peaks; all three peaks fall in $k\!\in\![5,7]$. At the $k\!=\!7$ default, every CI is strictly positive.}
\vspace{-0.5em}
\label{fig:threshold_k}
\end{figure}

\vspace{-0.2em}
\subsection{Intent-Aggregation Threshold $k$} \label{sec:abl-threshold-k}
We compare the production default $k\!=\!7$ against the at-least-one permissive baseline ($k\!=\!1$), which equates a single weak intent match with broad intent coverage.
We sweep $k \in [1, 15]$ on the same exhaustive event--candidate judgment matrix used in §\ref{sec:abl-intent} so the only varying quantity is the aggregation rule. Figure~\ref{fig:threshold_k} reports the Pearson correlation between the resulting MicroRel and CP, CR, and CVR measured per-(date, source) over a $30$-day production window across $2$ retrieval sources. All three correlations reach their maxima within $k\!\in\![5,7]$: P(CP) peaks at $0.43$ ($k\!=\!5$), and P(CR) and P(CVR) both peak at $k\!=\!7$. Beyond $k\!\geq\!8$, P(CP) and P(CR) decay as the candidate pass-rate drops below $\sim$$5\%$, while CVR retains most of its signal through $k\!=\!13$. At the $k\!=\!7$, all three $95\%$ bootstrap CIs are strictly positive: P(CP) $[+0.04, +0.63]$, P(CR) $[+0.02, +0.55]$, P(CVR) $[+0.29, +0.68]$. The intuition is that high-$k$ aggregation filters out single weak intent matches that are merely category-coincidental, requiring multi-intent support before a candidate counts as relevant, which aligns better with multi-touch purchase behavior. 

\vspace{-0.8em}
\section{System Efficiency} \label{sec:efficiency}
We benchmark four judge configurations on Vertex AI at 16-thread concurrency over $100$ production retrieval pairs (Table~\ref{tab:efficiency}). The fine-tuned Gemini-2.5-Flash we deploy reaches $751$\,ms p50, $1{,}117$\,ms p95, and $18$ QPS -- roughly $9\times$ faster than the teacher on p50 latency and on throughput. The fine-tuned Gemini-2.5-Flash-Lite variant offers a marginal further gain at $647$\,ms p50 and $20$ QPS. Most of the latency improvement therefore comes from fine-tuning rather than from picking a smaller backbone: prompt-only Flash still inherits the teacher's long-tail p95 ($8{,}807$\,ms), whereas the fine-tuned student compresses both p50 and p95 by an order of magnitude.

\vspace{-0.8em}
\section{Related Work}
\paragraph{Offline evaluation for recommendation and ads retrieval.}
Classical offline metrics for recommendation retrieval -- Recall@$K$, NDCG@$K$ over future-engagement labels, and conversion-attributed precision/recall~\cite{zhang2025optimizing} -- have well-known weaknesses in ads: severe selection bias from the exposure policy, heavy tail of low-engagement requests, and a lookback window that conflates retrieval quality with downstream ranker and budget effects. The offline-online evaluation gap in e-commerce recommendation has been documented at multiple scales~\cite{peska2020off}. Counterfactual estimators (IPS, doubly robust) reduce some of this bias~\cite{joachims2017accurately} but assume access to logged propensities that are rarely available at the retrieval stage of a multi-tower ad system. Our work is complementary: rather than debiasing engagement labels, we replace them with a content-grounded relevance label produced by an LLM judge, and report alignment with engagement-based offline metrics.

\paragraph{LLM judges for relevance.}
LLM-as-judge has moved from open-domain summarization~\cite{sun2023chatgpt} to e-commerce and recommender relevance labeling: LREF~\cite{tang2025lref} fine-tunes an LLM as a query-product relevance scorer with chain-of-thought, BloomIntent~\cite{choi2025bloomintent} performs intent-conditioned evaluation of SERP satisfaction, Wang et al.~\cite{wang2025semantic} progressively train an LLM-based semantic-ads retrieval encoder at Walmart, Penha et al.~\cite{penha2025cranfield} examine LLM-judge ranking agreement with humans in Cranfield-style movie and podcast recommendation, and large-scale studies show LLM labels broadly track human judgment in IR settings~\cite{rahmani2025judging, upadhyay2024large}. Two gaps remain. \emph{First}, these works rely on either an explicit query (search) or a Cranfield-style topic per request (movie/podcast collections); DPA retrieval has neither -- intent must be inferred from heterogeneous engagement signals, so we propose SID-grouped, event-stratified intent decomposition as the compression rule over the engagement history. \emph{Second}, prior LLM-judge studies validate against \textit{human assessors} at the pair or ranking level; we instead validate against \textit{online business KPIs} (CTR, CVR, conversion P/R) on the launch-decision critical path.

\begin{table}[t]
\centering
\small
\setlength{\tabcolsep}{4pt}
\renewcommand{\arraystretch}{1}
\caption{System efficiency of teacher and student judges. Latency is per-pair; QPS is aggregate throughput at the listed concurrency.}
\label{tab:efficiency}
\begin{tabular}{lccc}
\toprule
\textbf{Judge} & \textbf{p50 (ms)} & \textbf{p95 (ms)} & \textbf{QPS} \\
\midrule
Gemini-2.5-Pro (teacher, with guidel.)           & 6{,}963 & 10{,}840 & 2  \\
Gemini-2.5-Flash (prompt, with guidel.)          & 2{,}189 & 8{,}807  & 4  \\
\textbf{Gemini-2.5-Flash-ft (ours)}             & \textbf{751}   & \textbf{1{,}117}  & \textbf{18}  \\
Gemini-2.5-Flash-Lite-ft                        & 647   & 1{,}342  & 20  \\
\bottomrule
\end{tabular}
\vspace{-0.5em}
\end{table}

\vspace{-0.6em}
\section{Conclusion and Limitation} \label{sec:conclusion}
We present a query-free user--product relevance metric for offline evaluation of DPA retrieval. A user's engagement history is compressed into a small SID-grouped, event-stratified intent set, and an LLM judge scores each retrieved candidate against this set; an agentically-discovered guideline anchors a Gemini-2.5-Pro teacher, distilled into a Gemini-2.5-Flash student that matches the teacher at $\sim$$9\times$ lower latency. On 30 days of production DPA traffic, the metric ranks six retrieval sources consistently with conversion precision/recall and is directionally aligned with online CTR/CVR across four A/B tests; it serves as the offline gate for DPA model evaluation before online deployment.


\section{GenAI Usage Disclosure} \label{sec:GenAI_dis}
In accordance with the ACM Policy on the Use of Generative AI Tools and Technologies, we disclose that GenAI assistants (including ChatGPT and Claude) were used during the preparation of the manuscript for language editing and writing assistance, including improving grammar, clarity, readability, and consistency of terminology. We also used AI-assisted tools to help draft preliminary versions of certain figures and diagrams, which we subsequently reviewed, revised, and finalized. All research ideas, methodology, system design, experiments, analyses, results, and conclusions were developed and validated by the authors. All cited references were manually identified by the authors and verified against their original sources with the help of AI tools. Because this paper studies LLM-based relevance judge, the use of LLMs as components of the studied systems and experimental methodology is described separately in the relevant sections of the paper.

\bibliographystyle{ACM-Reference-Format}
\balance
\bibliography{sample-base}
\end{document}